\documentclass[floatfix, aps,prb,twocolumn, groupedaddress]{revtex4-2}
\usepackage{dcolumn}
\usepackage{bm}
\usepackage{amssymb}
\usepackage{amsthm}
\usepackage{physics}
\usepackage{color}
\usepackage[version=4]{mhchem}
\usepackage{graphicx}
\usepackage{natbib}
\usepackage{subfig}
\usepackage{booktabs}
\usepackage{multirow}
\usepackage[T1]{fontenc}

\newcommand{\degC}{$^\circ$C}

\newcommand{\TC}{T_{\rm C}}
\newcommand{\Tcp}{T_{\rm cp}}

\newcommand{\TCC}{T^*}
\newcommand{\TCW}{\theta_{\rm W}}
\newcommand{\peff}{p_{\rm eff}}
\newcommand{\uB}{\mu_{\rm B}}
\newcommand{\kB}{k_{\rm B}}
\newcommand{\Msat}{M_{\rm sat}}
\newcommand{\Mema}{M_{\rm sat}^{\rm EMA}}

\newcommand{\LIC}{\ce{LuInCo_4}}

\newcommand{\YMC}{\ce{YMgCo_4}}
\newcommand{\RIC}{\ce{$R$InCo_4}}
\newcommand{\GIC}{\ce{GdInCo_4}}
\newcommand{\TbIC}{\ce{TbInCo_4}}
\newcommand{\DIC}{\ce{DyInCo_4}}
\newcommand{\HIC}{\ce{HoInCo_4}}
\newcommand{\EIC}{\ce{ErInCo_4}}
\newcommand{\TIC}{\ce{TmInCo_4}}
\newcommand{\YbIC}{\ce{YbInCo_4}}
\newcommand{\RMC}{\ce{$R$MgCo_4}}
\newcommand{\RAC}{\ce{$RA$Co_4}}
\newcommand{\RC}{\ce{$R$Co_2}}
\newcommand{\YC}{\ce{YCo_2}}
\newcommand{\LuC}{\ce{LuCo_2}}
\newcommand{\ErC}{\ce{ErCo_2}}
\newcommand{\HoC}{\ce{HoCo_2}}

\newcommand{\DyC}{\ce{DyCo_2}}
\newcommand{\NIC}{\ce{NdCo_2}}

\newcommand{\Rion}{\ce{$R$^{3+}}}

\begin{document}


\title{Novel family of near-room-temperature compensated itinerant pyrochlore ferrimagnets, {\RIC} ($R$ = Dy--Tm)}


\author{Taiki Shiotani}
\email[]{Corresponding author: shiotani.taiki.48e@st.kyoto-u.ac.jp}
\affiliation{Department of Materials Science and Engineering, Kyoto University, Kyoto 606-8501, Japan}

\author{Takeshi Waki}
\affiliation{Department of Materials Science and Engineering, Kyoto University, Kyoto 606-8501, Japan}

\author{Yoshikazu Tabata}
\affiliation{Department of Materials Science and Engineering, Kyoto University, Kyoto 606-8501, Japan}

\author{Istv\'{a}n K\'{e}zsm\'{a}rki}
\affiliation{Experimental Physics 5, Center for Electronic Correlations and Magnetism, Institute of Physics, University of Augsburg, D-86159 Augsburg, Germany}

\author{Hiroyuki Nakamura}
\affiliation{Department of Materials Science and Engineering, Kyoto University, Kyoto 606-8501, Japan}

\date{\today}

\begin{abstract}
We successfully synthesized single crystals of a series of C15b Laves phase compounds, {\RIC} ($R$ = Dy--Tm), with Co-pyrochlore and $R$-fcc sublattices, and systematically studied their magnetic properties via magnetometry measurements.
These itinerant cubic compounds, with Curie temperatures above room temperature, show compensated ferrimagnetism featuring an antiferromagnetic coupling between the two sublattices. From this series, {\DIC} exhibits the highest $\TC\ (= 368$\ K) and a near-room-temperature compensation point $\Tcp\ (= 295$\ K).
$\TC$ does not change drastically with the $R$ atom, whereas $\Tcp$ depends on the de Gennes factor of {\Rion}. Another magnetization anomaly is observed in all the compounds at low temperatures, which may be indicative of changes in the lattice or magnetic structure.
The easy axis the ferrimagnetic moment of {\DIC}, {\EIC}, and {\TIC} is found at $T=5$\ K to be along the [001], [111] and [110] directions, respectively. However, the simple easy-axis or easy-plane ferrimagnetic picture cannot be applied to {\HIC}. 
These observations suggest that the $R$ sublattice determines magnetic anisotropy and compensation, while the Co sublattice plays a role in strong magnetic ordering. The high Curie temperature, together with the magnetization compensation point near room temperature, renders these itinerant pyrochlore magnets interesting for spintronic applications.
\end{abstract}


\maketitle

\section{Introduction}
\begin{figure}[h]
  \centering
  \includegraphics[keepaspectratio, width=1\columnwidth]{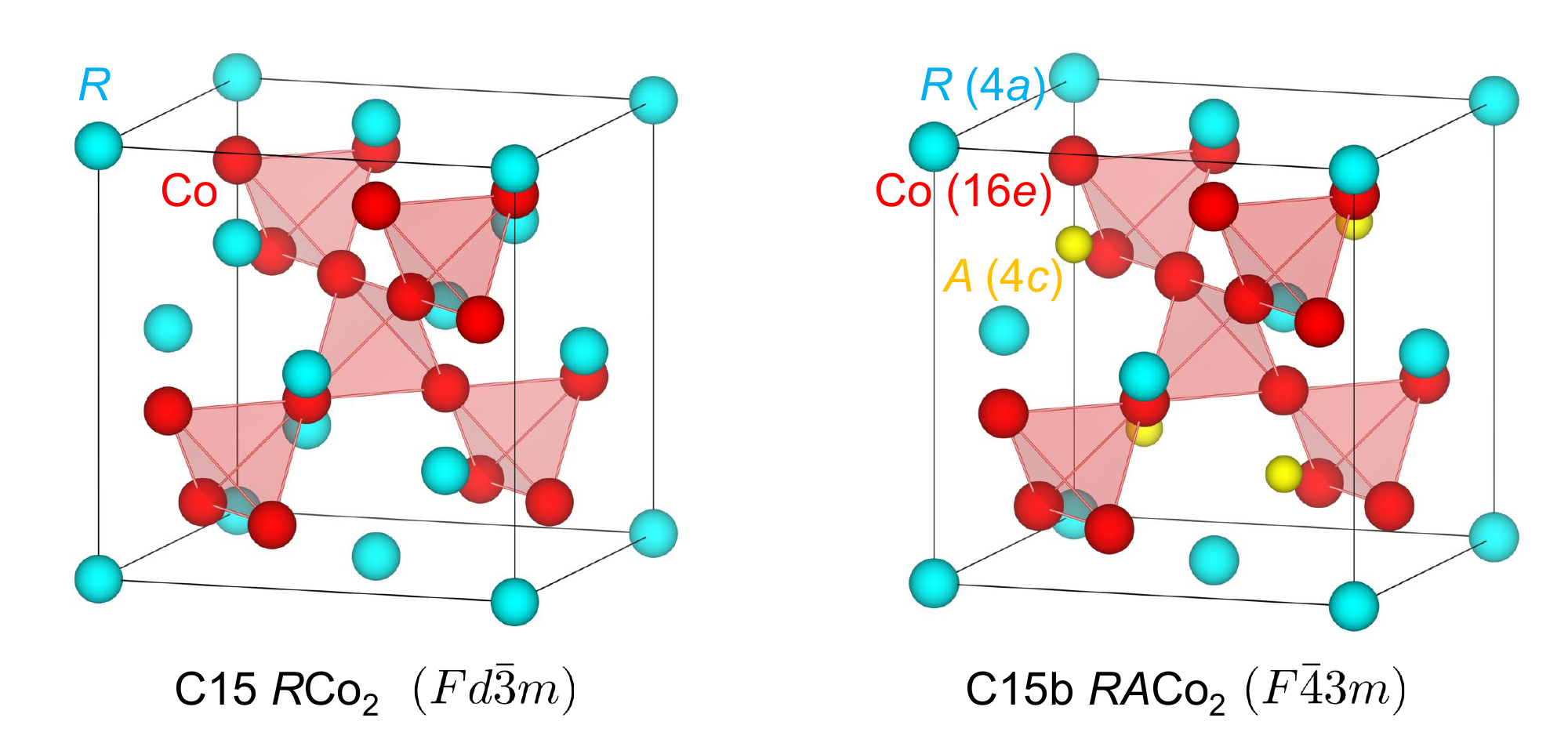}
  \caption{Crystal structures of the C15 and C15b Laves phase compounds.}
  \label{cryst}
\end{figure}
\indent 
In itinerant magnets, the magnetic properties as well as magneto-transport phenomena can be strongly influenced by the geometry of the underlying crystal lattice. Metallic magnets with a pyrochlore lattice, a network of corner-sharing tetrahedra, are attracting attention due to their exotic magnetic, magneto-optical and magneto-transport phenomena emerging therein.
{\ce{YMn_2}} is a well-known example; it exhibits helical magnetic ordering with an extremely long period due to the geometrical frustration of the antiferromagnetic interaction on the Mn-pyrochlore sublattice \cite{YMheli}.
Frustration in itinerant electron systems can also lead to spin-liquid behavior in materials such as Sc-substituted {\ce{YMn_2}} and {\ce{LiV_2O_4}} \cite{YMspl, LVO}. 
Recently, nontrivial band structures in pyrochlore metals have attracted attention.
Lattice frustration causes destructive interference in electron motion, leading to a three-dimensional flat bands \cite{flata, flatb}, which have been discovered experimentally in {\ce{CaNi_2}} and {\ce{CuV_2S_4}} \cite{CaNi, CuVS}. 
Furthermore, the lattice has been reported to host a four-fold degenerate Dirac point at the high-symmetry point of the Brillouin zone, which is protected by the crystalline nonsymmorphic symmetry \cite{diraca, diracb, CeRu}.
Moreover, even ferromagnetic pyrochlore magnets can exhibit unconventional magneto-transport and magneto-optical phenomena, like the scalar-spin-chirality induced anomalous Hall effect and its optical analogue, the magneto-optical Kerr effect \cite{iko, ikt, ikthr, ikf, ikfive}
\\
\indent The C15 Laves phase, {\RC} ($R$: rare-earth element), is a family of Co pyrochlore metals, in which the $R$ atom forms a diamond sublattice (Fig.\ \ref{cryst}). Despite the high density of states of the Co--3$d$ bands just below the Fermi level, these bands cannot be magnetically polarized without the molecular field from a magnetic $R$ sublattice \cite{RC}. Consequently, {\YC} and {\LuC} are exchange-enhanced Pauli paramagnets with high magnetic susceptibility \cite{Ymeta, Lumeta, lemaire, burzo}. Cases involving magnetic lighter $R$ elements are known to be ferromagnetic, while cases involving heavier $R$ elements are ferrimagnetic, with the $R$ and Co sublattices aligned antiparallelly \cite{ferroRC}.
The detailed characteristics of the ferrimagnetic {\RC} are sensitive to $R$; the highest Curie temperature, $\TC = 405$\ K, is found in {\ce{GdCo_2}}, but it decreases rapidly as the atomic number of $R$ moves away from Gd \cite{RCTC}.
Ferrimagnetic ordering is accompanied by lattice distortion from the cubic structure, and large spin anisotropy \cite{strDy, elasticLaves, LPdistortTm, elasticRC,Holevel, PrTbEraniso, CFE,Tbaniso, Tmaniso}. Lowered symmetry also depends on $R$, and further symmetry reduction has been observed in {\HoC} and {\ce{NdCo_2}} \cite{Ndnd, Hopm}.
\\
\indent Recently, a new series of cubic Laves phase compounds, {\RAC} ($A$\ =\ Mg and In), with a site-ordered C15b structure (Fig.\ \ref{cryst}), was reported for its synthesis and magnetic properties \cite{QQYMC, VVYMC, YMC, CeMC, TbMC, pottgen,LIC,RICsyn}. In $\RMC$, the Co atoms occupy Wyckoff position 16$e$ ($x$, $x$, $x$), the $R$ atoms 4$a$ (0, 0, 0), and the $A$ atoms 4$c$ ($\frac{1}{4}$, $\frac{1}{4}$, $\frac{1}{4}$) in the $F\bar{4}3m$ space group. Interestingly, {\RMC} has two frustrated sublattices: a breathing but nearly ideal pyrochlore sublattice of Co sites and a face-centered cubic (fcc) sublattice of $R$, leading to versatile quantum phases through the interplay of $R$-4$f$ and Co-3$d$ electrons.
Unlike the C15 Laves phase compounds, {\YMC} and {\LIC}, with nonmagnetic $R$, show strong ferromagnetism with Curie temperatures of $\TC = 405$\ K and 306\ K, respectively \cite{YMC,LIC}. DFT calculations suggest that the spontaneous spin polarization is induced by Co--3$d$ flat bands near the Fermi level.
{\LIC}'s unusual ferromagnetic nature is also intriguing; an anisotropic ferromagnetic state with an easy magnetization axis in the $\langle 001 \rangle$ direction emerges well below $\TC$ at 100 K, below which a metamagnetic transition occurs in low magnetic fields applied along the hard [111] direction. 
The observed flat bands and magnetic anomaly may indicate the importance of the Co-pyrochlore lattice geometry. Therefore, introducing a magnetic $R$ atom could be useful for tuning the magnetic properties and searching for new magnetic phases in magnetic pyrochlore metals. This would also reveal differences between the fcc and diamond lattices concerning the role of the coupling between $R$-4$f$ electrons and Co-3$d$ electrons. However, only the magnetic properties of {\ce{TbMgCo_4}} and {\YbIC} have been reported so far, except for the nonmagnetic $R$ cases \cite{TbMC, tsujii}.
\\
\indent In this context, we focus on {\RIC} with magnetic $R$ and report on the synthesis of single crystals for $R=$\ Dy--Tm. Based on magnetization measurements, we discuss the magnetic properties of the {\RIC} compounds and report for the first time that they exhibit ferrimagnetic transitions at temperatures higher than room temperature, with compensation temperatures that vary sensitively with $R$. Our analysis of the orientation-dependent isothermal magnetization suggests antiferromagnetic coupling between the $R$ and Co sublattices and strong magnetic anisotropy with spins aligned in the [001], [111] and [110] directions for {\DIC}, {\EIC} and, {\TIC}, respectively, while {\HIC} may posses a more complicated ferrimagnetic structure. 
In these compounds, we also report another magnetization anomaly at a lower temperature, which hints at the presence of a spin reorientation transition accompanied by lattice distortion.
\\
\section{Experiments}
The synthesis of {\RIC} polycrystals has only been reported for $R =$\ Dy--Lu \cite{RICsyn, tsujii}. 
Single crystals of {\RIC} were grown using the self-flux method described in our previous report on {\LIC} \cite{LIC}. We began with $R$ ingots (Rare Metallic, 99.9\% purity for Tb, Dy and Yb, and Johnson Matthey, 99.9\% purity for Gd, Ho--Tm), In shots (Rare Metallic, 99.99\% purity), and Co flakes (Rare Metallic, 99.9\% purity).
The mixture with a composition of {\ce{$R$Co_2In_{2.15}}} was arc-melted in an argon atmosphere, followed by grinding. The resulting mixture was placed in a boron nitride (BN) crucible and sealed in an evacuated quartz tube. The ampoule was initially heated up to 1200\ \degC\ over 3 h, held at that temperature for 0.5 h, cooled to 1150 \degC\ over 8 h, cooled slowly to 975\ \degC\ over 58h, and then centrifuged to separate the single crystals from the flux. Octahedral or truncated triangular crystals with \{111\} faces were obtained for $R$ = Dy--Tm.
No single crystals were available for {\GIC} and {\TbIC}. This method could not be applied to {\YbIC} because of the high vapor pressure of Yb.
The flux and impurity phases remaining on the surface were removed by immersion in dilute hydrochloric acid.
\\
\indent The grown crystals were characterized using x-ray diffraction (XRD) measurements with Cu\ $K_{\alpha_1}$ radiation with X’Pert PRO Alpha-1 (PANalytical). The Rietveld refinement was performed using Rietan-FP \cite{FP}. To this end, powder was prepared by crushing single crystals.
Magnetization measurements were performed using a superconducting quantum interference device magnetometer (MPMS-XL7, Quantum Design) in the temperature range of 5--350\ K and under magnetic fields up to 70 kOe. To determine the transition temperature, data were collected in the temperature range of 5--400 K along a specific crystal orientation using MPMS3.
Octahedral crystals were cut and used. The magnetic field was corrected by subtracting the demagnetizing field as $H = H_{\rm ext}- NM$, where $N$ is the demagnetizing factor. Each $N$ value was calculated using the representation from the literature \cite{demagprism} under the assumption that the sample shape was a rectangular prism. Note that this approximation introduces error in $N$. 
\\
\section{Results}
\subsection{Crystal structure}
\begin{figure}[ht]
    \centering
    \includegraphics[keepaspectratio, width=1\columnwidth]{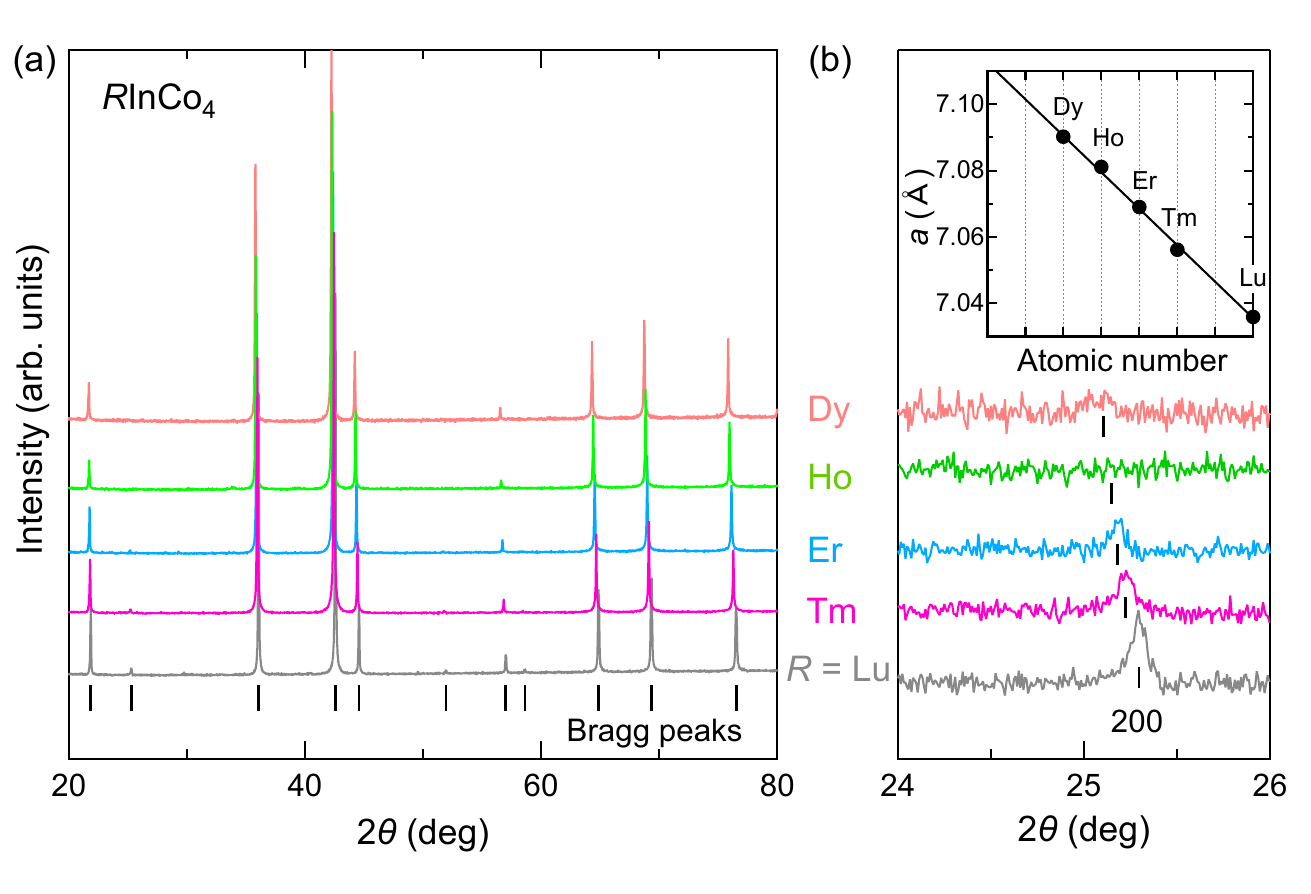} 
    \caption{(a) Powder XRD profiles of {\RIC} at room temperature.   (b) Profiles in the extended $2\theta$ range, where the 200 superstructure peak is observed. The inset shows the lattice parameter as a function of the $R$-atomic number. {\LIC} data were taken from to our previous report \cite{LIC}.}
    \label{xrd}
\end{figure}
\begin{table}[ht]
\centering
\caption{Results of the Rietveld refinement and the ratio of structure factors, $F_{200}/F_{111}$.}
\begin{ruledtabular}
\begin{tabular}{lccccc}
$R$ & $a$\ (\AA) & $x$ & $g_{4a}(R)$ & $g_{4c}$(In) & $F_{200}/F_{111}$ \\
\hline
Dy & 7.09038(1) & 0.6228(6) & 0.73(3) & 0.76(2) & 0.054  \\
Ho & 7.08108(1) & 0.6285(1) & - & - & 0.006  \\
Er & 7.06896(1) & 0.6245(2) & 1.01(2) & 0.97(2) & 0.062  \\
Tm & 7.05604(1) & 0.6234(3) & 0.88(1) & 0.96(2) & 0.086  \\
Lu \cite{LIC} & 7.03586 & 0.6252 & 0.99 & 0.87 & 0.120  \\
\end{tabular}
\label{fp}
\end{ruledtabular}
\end{table}
Figure\ \ref{xrd}(a) shows the powder XRD profiles of {\RIC} for $R\ =\ $ Dy--Tm at room temperature. They can all indicate that these compounds crystallizes in a single phase. The 200 superstructure peak, which is characteristic of the C15b site-ordered structure with the $F\bar{4}3m$ space group was observed for all $R$ except $R\ =\ $Ho (Fig.\ \ref{xrd}(b)). This peak is difficult to discern, especially for {\HIC}, when using Cu\ $K_{\alpha_1}$ radiation, since its structure factor, $F_{200}$, is relatively small compared to other reflections. For example, under the assumption of perfect ordering, $F_{200}/F_{111}$ is expected to be approximately 0.006 for {\HIC}, whereas 0.05--0.12 for $R=$\ Dy and Er--Lu (Table\ \ref{fp}).
We performed structure refinement using the Rietveld method with Rietan-FP \cite{FP}. 
The refined parameters are shown in Table\ \ref{fp}. The lattice parameter decreases linearly with the atomic number of $R$ as shown in the inset in Fig.\ \ref{xrd}(b), which is related to dominant lanthanide contraction.
The calculated atomic coordinates at the Co\ $16e\ (x, x, x)$ are all close to the coordinates of the non-breathing pyrochlore lattice ($x=0.625$), similar to {\LIC} \cite{LIC}.
The site occupancy of $R$/In at the 4$a$ site and In/$R$ at the 4$c$ site were $g_{4a}({\ce{$R$}}) = 1.01$(2) and $g_{4c}({\ce{In}}) = 0.97$(2) for $R=\ $Er and $g_{4a}({\ce{$R$}}) = 0.88$(1) and $g_{4c}({\ce{In}}) =  0.96$(2) for $R=\ $Tm, which verifies the nearly perfect atomic ordering of $R$ and In. 
Conversely, we estimated that {\DIC} has non-negligible disorder ($g_{4a}({\ce{$R$}}) = 0.73$(3) and $g_{4c}({\ce{In}}) = 0.76$(2)) and {\HIC} has site occupancies much larger than one and negative, which are physically unreasonable. Since the small $F_{200}$ values can affect these refined values, a powder diffraction study using different radiation is needed for more accurate refinement.  
The temperature at which the measurements were taken ($T \approx 300$\ K) is lower than the Curie temperatures of {\RIC} (see Fig.\ \ref{MT}), where no signs of symmetry lowering from the cubic structure were observed within the resolution.
This contrasts with C15 ferrimagnets, {\RC}, with the significant lattice distortion at $\TC$ \cite{strDy, elasticLaves, LPdistortTm, elasticRC}. This difference, that is surprising in the light of the strong spin-orbit coupling specific to rare earth elements, can be attributed to the magnetic sublattice that dominates the magnetic ordering, as discussed in the section on the temperature dependence of magnetization. 
\\
\subsection{Temperature dependence of magnetization}
\begin{figure*}[ht]
    \includegraphics[keepaspectratio, width=2\columnwidth]{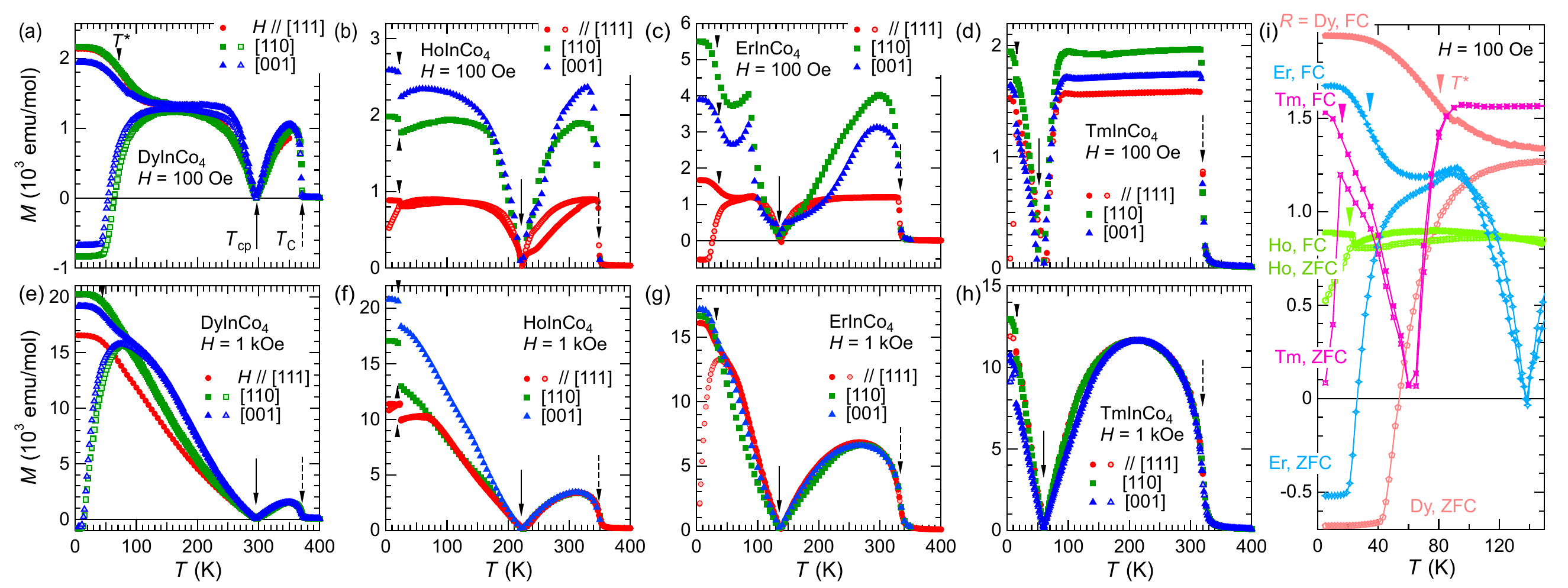}
    \caption{Temperature dependence of the magnetization of {\RIC} ($R$ = Dy--Tm) at (a)--(d) $H=100$\ Oe and (e)--(h) $H=1$\ kOe. Dark and light markers represent data obtained under field-cooled and zero-field-cooled conditions, respectively. (i) data across the magnetization anomaly at $\TCC$. A field of $H = 100$\ Oe was applied in the [001] direction for $R$ = Dy and in the [111] direction for $R$ = Ho--Tm.}
    \label{MT}
\end{figure*}
    \begin{figure}[ht]
    \centering
    \includegraphics[keepaspectratio, width=0.8\columnwidth]{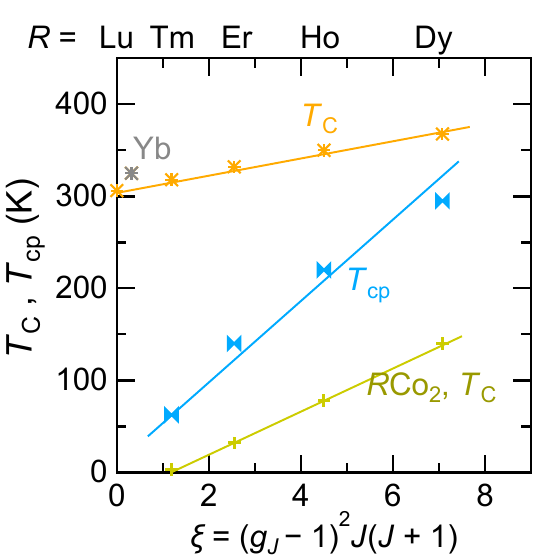}
    \caption{Curie and compensation temperatures as a function of the de Gennes factor, $\xi$. Data of {\LIC}, {\YbIC}, and {\RC} were taken from Ref. \cite{LIC}, \cite{tsujii}, and \cite{RCTC}, respectively. Lines are guides to the eye.}
    \label{TC}
\end{figure}
\begin{figure}[ht]
    \centering
    \includegraphics[keepaspectratio, width=0.8\columnwidth]{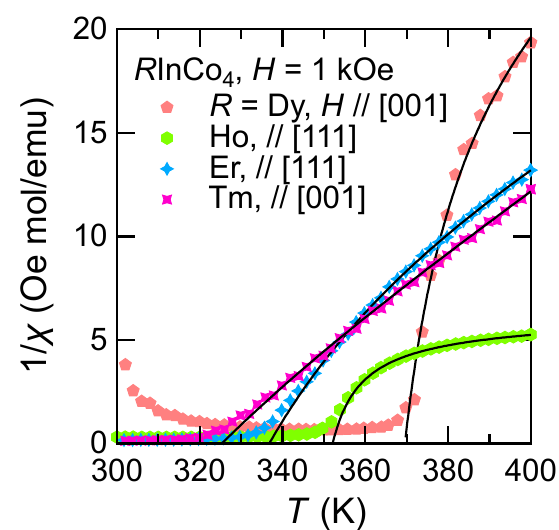}
    \caption{The inverse susceptibility of {\RIC}. The solid curves represent the fit to the Eq.\ \ref{Neel}.}
    \label{inv}
\end{figure}
\begin{table}[ht]
\centering
\caption{Transition temperatures ($\TC$, $\Tcp$) and expected effective moment ($\peff$). Results of fitting the susceptibility to Eq.\ \ref{Neel} and the modified Curie-Weiss law (Mod. CW).}
\begin{ruledtabular}
\begin{tabular}{lrcccc}
&$R=$ & Dy & Ho & Er & Tm \\
\hline
&$\TC$\ (K)  & 368 & 350 & 332 & 318 \\
&$\Tcp$\ (K) & 295 & 220 & 140 & 62.5 \\
&$\peff$(calc) \ ($\uB$) & 14.28  & 14.25  & 13.50  & 12.15  \\
\hline
Eq.\ \ref{Neel}&$\peff$\ ($\uB$) & 13.07 & 14.24 & 14.42 & 14.38 \\
&$\theta$\ (K) & 360 & 333 & 294 & 245 \\
&$T_a$\ (K) & 30 & 41 & 14 & 0 \\
&$\sigma$\ (emu/mol$\cdot$Oe) & 173 & 304 & 455 & 1014 \\
\hline
Mod. CW&$\peff$\ ($\uB$) & 1.98 & 3.38 & 5.31 & 7.08 \\
&$\TCW$\ (K) & 371 & 351 & 334 & 322 \\
\end{tabular}
\label{CW}
\end{ruledtabular}
\end{table}

Figure\ \ref{MT} shows the temperature dependence of the magnetization of {\RIC} for $R$ = Dy--Tm in external fields of $H=100$\ Oe and 1\ kOe. Overall, the {\RIC} compounds show compensated ferrimagnetism accompanied by magnetization anomalies at low temperatures.
{\DIC} exhibits the highest transition temperature ($\TC = 368$\ K) and the compensation temperature ($\Tcp = 295$\ K). 
Figure\ \ref{TC} shows $\TC$ and $\Tcp$ as a function of the de Gennes factor, $\xi = (g_J-1)^2J(J+1)$, where $J$ is the total angular momentum quantum number and $g_J$ is the Land\'{e} $g$-factor of {\ce{$R$^{3+}}}. $\TC$ and $\Tcp$ both decrease with decreasing $\xi$, i.e., with heavier $R$. $\Tcp$ is almost proportional to $\xi$. This is consistent with the observation that {\LIC} ($\xi = 0$) has no compensation point. The slope is similar to that of $\TC$ for {\RC} \cite{RCTC}.  
Conversely, $\TC$ behaves linearly, albeit with a small slope and a large intercept, that is 306 K for {\LIC} \cite{LIC}. It has been suggested that the antiferromagnetic intersublattice coupling depends on the $R$, while the spontaneous magnetic ordering is dominated by the Co sublattice with highly polarized 3$d$ bands \cite{LIC}.
Note that the $\TC$ of polycrystalline {\YbIC} was reported to be $\TC = 325$\ K \cite{tsujii}, and the deviation from the linear $\TC$--$\xi$ relation may be due to slight differences in the sample quality or compositions depending on whether the sample is single- or polycrystalline. 
This nature of ferrimagnetism differs from that of {\RC}, where the 3$d$ bands cannot be polarized without the molecular field from the magnetic $R$ atom. The small variation in the $\TC$ of {\RIC} is more likely to be due to the effect of the volume expansion (Fig.\ \ref{xrd}).
\\
\indent The inverse susceptibility is shown in Fig.\ \ref{inv}. Above $\TC$, only {\TIC} shows a linear increase and the others show a hyperbolic increase with a negative curvature.
In general, the inverse susceptibility of ferrimagnets follows the Curie-Weiss law with a negative $\TCW$ at high temperature and deviates significantly towards smaller values near the $\TC$, depending on the intrasublattice and intersublattice molecular field coefficients \cite{NeelCW, MnONeelCW}. The expression for the hyperbolic inverse susceptibility in ferrimagnets is given as\\
\begin{equation}\label{Neel}
\frac{1}{\chi - \chi_0 } = \frac{T}{C} + \frac{1}{\chi_a} + \frac{\sigma}{T - \theta},
\end{equation}
where $C$ is the Curie constant, and $\chi_a$, $\sigma$, and $\theta$ consist of the Curie constants of each magnetic sublattice and the molecular field coefficients. The term $\chi_0$ consists of other temperature-independent contributions, such as the diamagnetic, Pauli paramagnetic and van Vleck paramagnetic susceptibility. 
We fitted the Eq.\ \ref{Neel} to the inverse susceptibility and the results are shown in Table\ \ref{CW}.
The effective paramagnetic moment, $\peff$, is estimated experimentally using the relation $C = \uB \peff ^2/3\kB$. Since we are considering $R$ and Co magnetic atoms, $\peff$ is expected to be $\peff =  \sqrt{{\peff (R^{3+}) }^2 + 4  {\peff ({\ce{Co}})}^2}$, where $\peff (R^{3+}) = g_JJ(J+1)\uB$ and $ \peff ({\ce{Co}}) \approx 3.22\ \uB$, also obtained from literature data on {\LIC} \cite{LIC}.
The asymptotic Curie temperature, $T_a$, in the high-temperature range, where the last term of Eq.\ \ref{Neel} is negligible, can be obtained from $T_a = C/\chi_a$.
For {\DIC} and {\HIC}, the experimental $\peff$ and the paramagnetic Curie point $\theta$ are close to the expected {$\peff$} and the ferrimagnetic transition point $\TC$, respectively. However, there is a significant discrepancy between $\TC$ and $\theta$ for {\EIC} and {\TIC}. 
Data in the high-temperature regime, where the asymptotic behavior appears, is needed to achieve a better fit and to analyze the coefficients of the intersublattice and intrasublattice molecular fields. 
We also attempted to fit the data to the modified Curie-Weiss law $\chi - \chi_0 = C/(T - \TCW)$, where $\TCW$ is the Weiss temperature. The fitting results are shown in Table\ \ref{CW}. 
Although the $\TCW$ is close to the $\TC$, the experimental $\peff$ is found to be substantially underestimated compared to the calculated value. These results suggest that the model with Eq.\ \ref{Neel}, which applies the molecular field theory to ferrimagnetism, is more appropriate. 
\\
\indent In addition to the compensated ferrimagnetism, Figure\ \ref{MT} shows that {\RIC} exhibits complex magnetization behavior at low temperatures. When {\DIC} is cooled from $T \approx 150$\ K in a field of $H = 100$\ Oe, the field-cooled (FC) magnetization increases, while the zero-field-cooled (ZFC) magnetization decreases. 
This bifurcation becomes more pronounced at $\TCC \approx 80$\ K and shifts to lower temperatures at a higher field of $H= 1$\ kOe. Note that the negative ZFC magnetization is due to incomplete demagnetization.   
{\HIC} shows a step-like increase in the FC magnetization and a rapid decrease in the ZFC magnetization when cooled down to $\TCC = 22$\ K. Although the bifurcation is suppressed at a higher field of $H= 1$\ kOe, the abrupt increase still persists.
{\EIC} exhibits two characteristic points. At $T = 85$\ K and $H = 100$\ Oe, the magnetization displays a cusp-like anomaly and begins to bifurcate between the FC and ZFC conditions. The bifurcation becomes more pronounced at $\TCC = 25$\ K, and both are suppressed by high fields, similarly to {\DIC}.
{\TIC} shows an anomaly similar to that of {\HIC} at the lower temperature of $\TCC = 15$\ K.
These types of anomalies have also been observed in C15 Laves phase compounds, {\RC}. {\HoC} and {\NIC} show a spin-reorientation transition accompanied by a structural change from the tetragonal to the orthorhombic phase at $\TCC = 15$\ K and 43 K, respectively \cite{Ndnd, Ndxrd, Hotsr, Hopm}, which has been attributed to a temperature-induced transition between the spin configurations of the {\ce{$R$^{3+}}} ion \cite{Holevel, CFE}.  
{\DyC} and {\ErC} have been reported to exhibit a magnetization anomaly at low temperatures and at low fields, which is similar to, but weaker than, that of {\DIC} and {\EIC}, respectively \cite{Dyano, Erano}. In these two cases, the relationships with the change in magnetic and crystal structures is unclear.  
While studying on the crystal structure and the temperature dependence of the magnetic anisotropy would provide useful insight into further understanding of the low-temperature magnetism of {\RIC}, the comprehensive investigation on the the microscopic origin of the anomalies at $\TCC$ is beyond the scope of this work.
The relatively small bifurcation of the ZFC and FC curves can be attributed to pinning of magnetic domain walls. The distinct bifurcation below $\TCC$ is considered to be associated with a drastic change in the magnetic anisotropy or emergent domain splitting between crystallographically different phases along with a structural change \cite{DWhys}. 
\\
\section{Isothermal magnetization}
\begin{figure*}[ht]
    \includegraphics[keepaspectratio, width=2\columnwidth]{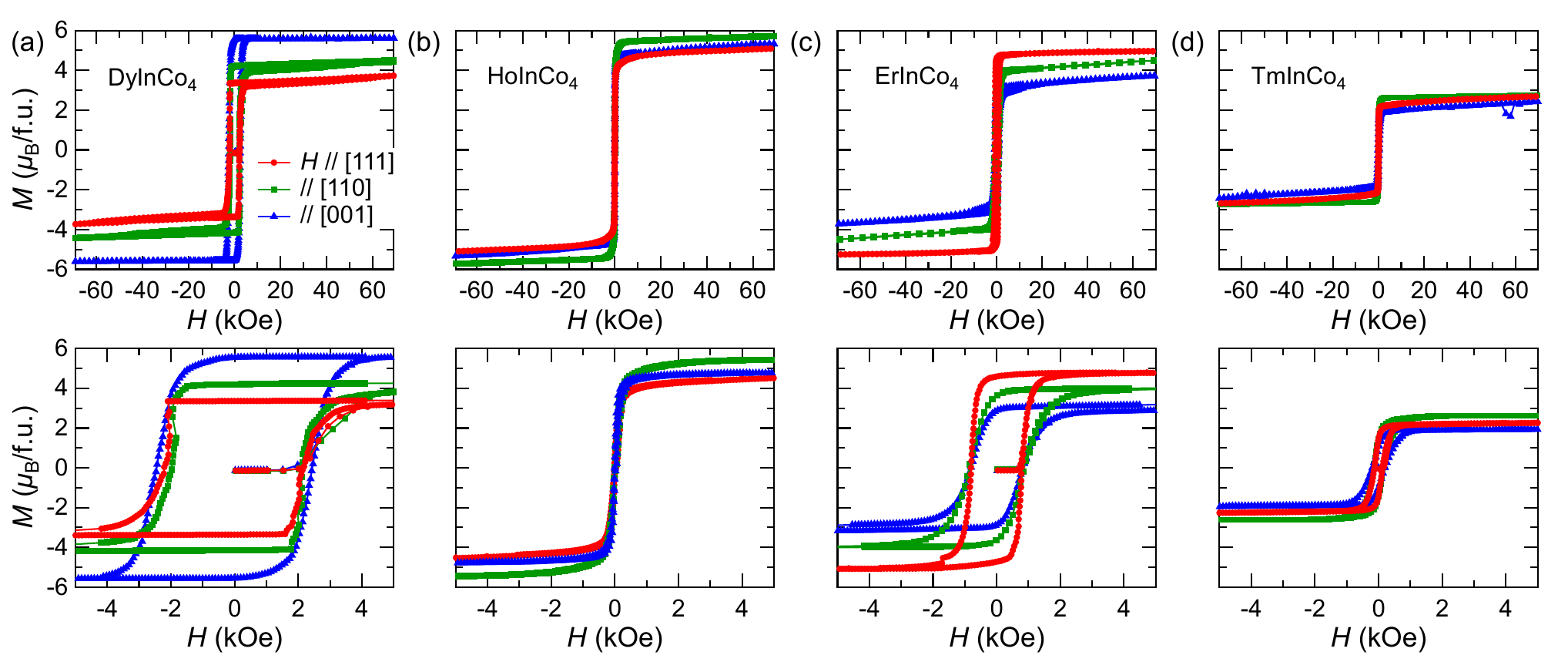}
    \caption{ (a)--(d) Field dependence of the magnetization of {\RIC} ($R$ = Dy--Tm) at $T=5$\ K. The lower figures show the data from the low-field region.}
    \label{mh}
\end{figure*}
\begin{table*}[ht]
\centering
\begin{ruledtabular}
\caption{Saturated magnetization of {\RIC} along the crystal principal axes, the calculated value along the easy-magnetization axis, $\Mema = \mu_{\Rion} - \Msat^{\LIC} $, the expected easy magnetization axis or plane, and the calculated $\Msat$ values using the Eq.\ \ref{Malpha}.}
\begin{tabular}{lccccc}
 & {\DIC} & \multicolumn{2}{c}{\HIC} & {\EIC} & {\TIC} \\
\hline
$\Msat ^{[001]}$ & 5.57 & \multicolumn{2}{c}{4.70} & 2.75 & 1.89 \\
$\Msat ^{[110]}$ & 4.04 & \multicolumn{2}{c}{5.44} & 3.93 & 2.62 \\
$\Msat ^{[111]}$ & 3.20 & \multicolumn{2}{c}{4.70} & 4.83 & 2.16 \\
Calculated $\Mema$ & 6.5 & \multicolumn{2}{c}{6.5} & 5.5 & 3.5 \\
\hline
EMA/EMP & $\langle001\rangle$ & \ \ \ $\langle110\rangle$ & (111)\ \ \  & $\langle111\rangle$ & $\langle110\rangle$ \\
Calculated $\Msat ^{[001]}$ & 5.57 & \ \ \  3.85 & 4.44\ \ \  & 2.79 & 1.85 \\
Calculated $\Msat ^{[110]}$ & 3.94 & \ \ \ 5.44 & 5.44\ \ \  & 3.94 & 2.62 \\
Calculated $\Msat ^{[111]}$ & 3.22 & \ \ \ 4.44 & 5.12\ \ \  & 4.83 & 2.14 \\
\end{tabular}
\label{msat}
\end{ruledtabular}
\end{table*}
Figure\ \ref{mh} shows the field dependence of the magnetization for the four {\RIC} compounds at $T=5$\ K. They all exhibit ferromagnetic behavior. Relatively large hysteresis is evident in all crystal directions for {\DIC} and {\EIC} whereas it is minimal in {\HIC} and {\TIC}, consistent with the trend of the thermal hysteresis (Fig.\ \ref{MT}).   
The saturated magnetization, $\Msat$, depends on the crystal direction, indicating that they have a significant anisotropy. The $\Msat$ value calculated from the linear extrapolation to zero field for each principal direction is shown in Table\ \ref{msat}.
We assume a cubic single-ion anisotropy of the $R$ moments so strong that the magnetization cannot easily rotate from the easy-magnetization axis (EMA) to the applied field directions, thus in the low-field regime the projection of the magnetization to the direction of the applied field is measured.
Given that $\Msat$ of {\LIC} was reported to be 3.43\ $\uB$/f.u., the $\Msat$ of {\RIC} along the EMA can be simply expressed as $\Mema = \mu_{\Rion} \pm \Msat^{\LIC} = g_JJ \pm 3.43$\ ($\uB$/f.u.). The maximum value measured in the three directions (Table\ \ref{msat}) is close to $g_JJ - 3.43$, which is calculated for antiparallel spin coupling between the $R$-fcc and Co-pyrochlore sublattices. This is consistent with the ferrimagnetic structure of {\ce{TbMgCo_4}} and {\RC} with heavy $R$ \cite{TbMC,ferroRC}. The deviation is attributed to the Co moment magnitude, which varies with the molecular field from the $R$ atom and itself.
When the field is applied in a direction not equivalent to the EMA, $\Msat$ can be represented as
\begin{equation}\label{Malpha}
\Msat = \Mema \cos{\alpha},
\end{equation}
where $\alpha$ is the angle between the field direction and the EMA.  
If the EMA of {\DIC} is the $\langle001\rangle$ direction, then the [110] and [111] field direction form angles of $\alpha = 45^\circ$ and $\alpha = 54.7^\circ$, respectively. Using these $\alpha$ and experimental $\Mema$ values, calculated $\Msat$ values are close to the experimental values (listed in Table\ \ref{msat}).
Similarly, for {\EIC}, the $\langle111\rangle$ easy axis forms an angle of $\alpha = 54.7^\circ$ with the [001] direction and $\alpha = 35.3^\circ$ with the [110] direction.
For {\TIC}, the $\langle110\rangle$ easy axis forms an angle of $\alpha = 45^\circ$ with the [001] direction and $\alpha = 35.3^\circ$ with the [111] direction. Using these $\alpha$ values, the Eq.\ \ref{Malpha} yields calculated $\Msat$ values which agree well with experimental data.
However, for {\HIC}, the $\langle110\rangle$-easy axis model shows discrepancies with the experimental data, particularly in the [001] direction.
We then considered the easy-magnetization plane (EMP) cases of the \{001\}, \{110\} and \{111\} planes. In the \{001\} and \{110\} easy plane cases, $\Msat ^{[001]}$ can be equal to $\Mema$, as the [001] field direction is in planes ($\alpha = 0^\circ$) such as (100) and $(110)$. However, this is not the case.
For the \{111\} easy plane case, the [001] direction forms a minimum angle of $\alpha = 35.3^\circ$ when the moment is aligned, for example, in the $[11\bar{2}]$ direction in the (111) plane. 
Similarly, [111] forms an angle of $\alpha = 19.5^\circ$ with the [112] direction in the $(11\bar{1})$ plane, and [110] forms an angle of $\alpha = 0^\circ$ with the [110] direction in the $(1\bar{1}1)$ plane. In this case, the calculated $\Msat ^{[001]}$ is closer to the experimental value, but $\Msat ^{[111]}$ is not.
Therefore, it is necessary to consider the more complex spin structures such as collinear ones tilted from the principal axis, as well as noncollinear ones due to the competition between single-ion anisotropies between the Ho and Co sublattices.
\\
\section{Conclusion}
We successfully synthesized single crystals of the C15b Laves phase compounds {\RIC}, where $R$ = Dy--Tm, with Co-pyrochlore and $R$-fcc sublattices. Magnetization measurements revealed that the {\RIC} compounds are compensated ferrimagnets between these two sublattices with $\TC$ above room temperature, which is higher than those of {\RC} by more than 150 K. Of these compounds, {\DIC} has the highest $\TC$ of 368\ K and $\Tcp$ of 295\ K, which is near room temperature.
High-temperature magnetic ordering is stabilized by the Co magnetic sublattice, while the compensation depends on the $R$ atom and its de Gennes factor. All of them exhibit another magnetic anomaly at low temperatures suggesting the possibility of the spin reorientation with significant lattice distortion.
{\DIC}, {\EIC}, and {\TIC} are estimated to have collinear ferrimagnetic structures with the moments aligned in the [001], [111] and [110] directions, respectively at $T=5$\ K. In these compounds the anisotropy observed in the saturation moment indicates the orientational confinement of the $R^{3+}$ spins in low fields, likely governed by the cubic magnetocrystalline anisotropy. Conversely, a complex ferrimagnetic structure is suggested in {\HIC}. 
Magnetic anisotropy is considered to be related to the lowering of crystal symmetry, which is governed by the $R$-4$f$ electrons. Understanding the relationship between the crystal and magnetic structures are highly desirable, for example by low-temperature X-ray diffraction, neutron diffraction, and nuclear magnetic resonance.
Our work suggests that the {\RIC} family exhibits a variety of the physical properties similar to those of magnetic pyrochlore metals. Additionally it has the potential to be new compensated ferrimagnets near room temperature for use in spintronics or magnetocaloric devices.

\begin{acknowledgments}
This work was supported by JSPS KAKENHI Grant No. JP24KJ1325. The authors thank L. Prodan for their help with magnetization measurements. 
\end{acknowledgments}

\bibliography{ref}

\end{document}